\documentclass[letterpaper,journal]{IEEEtran}

\usepackage{amsmath}
\usepackage{array}
\usepackage{cite}
\usepackage{csquotes}
\usepackage[pdftex]{graphicx}
\usepackage{microtype}

\usepackage{algorithmic}
\usepackage{url}
\usepackage{listings}
\usepackage[capitalise]{cleveref}
\usepackage{multirow}
\usepackage{verbatim}

\usepackage[acronym,nomain,nonumberlist,nopostdot]{glossaries}
\usepackage{upgreek}
\usepackage{booktabs}
\usepackage{arydshln}
\usepackage{geometry}
\usepackage{ulem}

\usepackage{url}
\usepackage{tikz}
\usetikzlibrary{calc}

\newcommand{\positiontextbox}[4][]{%
	\begin{tikzpicture}[remember picture,overlay]
		\node[inner sep=3pt, fill=yellow,align=left,draw,line width=1pt,#1] at ($(current page.north west) + (#2,-#3)$) {\parbox{.80\paperwidth}{#4}};
	\end{tikzpicture}%
}

\newif\ifshowtodos

\ifshowtodos
   \usepackage[colorinlistoftodos,prependcaption]{todonotes}
\else
   \usepackage[disable, colorinlistoftodos,prependcaption]{todonotes}
\fi
 
\usepackage{newfloat}
\makeatletter
  \if@todonotes@disabled%
    \newcommand{\listofdonetodos}[1][]{}
    \newcommand{\listofchecktodos}[1][]{}

    \newcommand\todocheck[3][]{}
  \else
    \DeclareFloatingEnvironment[fileext=tdod,listname={Todo resolved}]{donetodo}

    \DeclareFloatingEnvironment[fileext=tdodc,listname={Todo Check}]{checktodo}
    \newcommand{\todocheck}[3][color=gray]{\todo[nolist,#1!60,textcolor=black,bordercolor=yellow!100]{#2  \textbf{#3}}
    \addcontentsline{tdodc}{checktodo}{#2 \textbf{#3}}}
        
  \fi 
\makeatother

\newacronym{ADC}{ADC}{analog-to-digital converter}
\newacronym{ANN}{ANN}{artificial neural network}
\newacronym{AOA}{AOA}{angle of arrival}
\newacronym{AOD}{AOD}{angle of departure}
\newacronym{AOP}{AOP}{antenna on package}
\newacronym{AP}{AP}{access point}
\newacronym{BLE}{BLE}{Bluetooth/Bluetooth low energy}
\newacronym{CAGR}{CAGR}{compound annual growth rate}
\newacronym{CDF}{CDF}{cumulative distribution function}
\newacronym{CFAR}{CFAR}{constant false alarm rate}
\newacronym{CITIC}{CITIC}{Centre for Information and Communications Technology Research}
\newacronym{CIR}{CIR}{channel impulse response}
\newacronym{DAC}{DAC}{digital-to-analog converter}
\newacronym{DBSCAN}{DBSCAN}{density-based spatial clustering of applications with noise}
\newacronym{DMA}{DMA}{direct memory access}
\newacronym{DNN}{DNN}{deep neural network}
\newacronym{DSP}{DSP}{digital signal processor}
\newacronym{ECDF}{ECDF}{empirical cumulative distribution function}
\newacronym{EKF}{EKF}{extended Kalman filter}
\newacronym{FFT}{FFT}{fast Fourier transform}
\newacronym{FMCW}{FMCW}{frequency modulated continuous wave}
\newacronym{FoV}{FoV}{field of view}
\newacronym{FTM}{FTM}{fine time measurement}
\newacronym{GNSS}{GNSS}{global navigation satellite system}
\newacronym{GP}{GP}{Gaussian process}
\newacronym{GPIO}{GPIO}{general-purpose input/output}
\newacronym{HWA}{HWA}{radar hardware accelerator}
\newacronym{I2C}{I$^2$C}{inter-integrated circuit}
\newacronym{I2S}{I$^2$S}{inter-IC sound}
\newacronym{INS}{INS}{inertial navigation system}
\newacronym{IF}{IF}{intermediate frequency}
\newacronym{IOT}{IoT}{internet of things}
\newacronym{IR}{IR}{infrared}
\newacronym{LBS}{LBS}{location based services}
\newacronym{LCD}{LCD}{liquid-crystal display}
\newacronym{LDO}{LDO}{low-dropout regulator}
\newacronym{LED}{LED}{light-emitting diode}
\newacronym{LNA}{LNA}{low-noise amplifier}
\newacronym{LOS}{LoS}{line of sight}
\newacronym{MCU}{MCU}{microcontroller unit}
\newacronym{MMWAVE}{mmWave}{millimiter wave}
\newacronym{ML}{ML}{machine learning}
\newacronym{MQTT}{MQTT}{message queuing telemetry transport}
\newacronym{NLOS}{NLOS}{non line of sight}
\newacronym{NN}{NN}{neural network}
\newacronym{OPTICS}{OPTICS}{ordering points to identify the clustering structure}
\newacronym{PA}{PA}{power amplifier}
\newacronym{PWM}{PWM}{pulse-width modulation}
\newacronym{ReLU}{ReLU}{rectified linear unit}
\newacronym{RF}{RF}{radio frequency}
\newacronym{RFID}{RFID}{radio frequency identification}
\newacronym{ROM}{ROM}{read-only memory}
\newacronym{ROS}{ROS}{Robot Operating System}
\newacronym{RSS}{RSS}{received signal strength}
\newacronym{RSSI}{RSSI}{received signal strength indicator}
\newacronym{RTT}{RTT}{round trip time}
\newacronym{SNR}{SNR}{signal-to-noise ratio}
\newacronym{SRAM}{SRAM}{static random-access memory}
\newacronym{SPI}{SPI}{serial peripheral interface}
\newacronym{SOC}{SOC}{system on a chip}
\newacronym{SVM}{SVM}{support vector machine}
\newacronym{TDOA}{TDOA}{time difference of arrival}
\newacronym{TLV}{TLV}{type length value}
\newacronym{TOA}{TOA}{time of arrival}
\newacronym{TOF}{TOF}{time of flight}
\newacronym{UART}{UART}{universal asynchronous receiver-transmitter}
\newacronym{UWB}{UWB}{ultra wideband}

\begin{document}
%
\title{An IoT System for Smart Building Combining Multiple mmWave FMCW Radars Applied to People Counting}

\author{Valentín Barral,
Tomás Domínguez-Bolaño,
Carlos J. Escudero,~\IEEEmembership{Senior Member,~IEEE}, and
José A. García-Naya,~\IEEEmembership{Senior Member,~IEEE}
\thanks{Manuscript received XX XXXX 2024; revised XX XXXX 2024
and XX XXXX 2024; accepted XX XXXX 2024. Date of publication
XX XXXX 2024; date of current version XX XXXX 2024. This work was supported in part by grants PID2022-137099NB-C42 (MADDIE) and TED2021-130240B-I00 (IVRY) funded by MCIN/AEI/10.13039/501100011033; in part by the European Union NextGenerationEU/PRTR; and funding for open access charge: Universidade da Coruña/CISUG. \textit{Corresponding author: Valentín Barral Vales.}

Special thanks to Ángel Carro Lagoa for his assistance in person tracking using images.\cite{CARROLAGOA2023100940}

The authors are with the Universidade da Coruña (University of A Coruña), CITIC Research Center, 15071 A Coruña, Spain. (e-mail: \mbox{valentin.barral@udc.es}, \mbox{tomas.bolano@udc.es}, \mbox{escudero@udc.es}, and \mbox{jagarcia@udc.es})
}}

\markboth{IEEE Internet Of Things Journal,~Vol.~XX, No.~XX, XXX~2024}{}

\maketitle

\listoftodos
{\let\clearpage\relax \listofdonetodos}
{\let\clearpage\relax \listofchecktodos}
\begin{abstract}
In contemporary society, the pressing challenge of preserving user privacy clashes with the imperative for smart buildings to efficiently manage their resources, particularly in the context of occupancy monitoring for optimized energy utilization. 
This paper delves into the application of \gls{MMWAVE} \gls{FMCW} radar technology for occupancy monitoring.  \gls{MMWAVE} \gls{FMCW} radar,
unlike conventional methods that often require the use of identifiable tags or involve image analysis, operates without the need for such identifiers, mitigating privacy concerns. 
However, challenges arise when attempting to cover extensive indoor spaces due to the limited range of individual mmWave \gls{FMCW} radar devices. The present work proposes the use of a flexible software architecture to integrate the measurements of several mmWave \gls{FMCW} radar devices, so that the whole behaves as a single sensor.
To validate the proposal, an example of use in a real environment in an indoor space monitored with three mmWave \gls{FMCW} radar devices is also presented. The example details the whole process, from the physical installation of the devices to the use of the different software modules that allow the integration of the data into a common \gls{IOT} management platform such as Home Assistant. All the elements, from the measurements captured during the test to the different software implementations, are shared publicly with the scientific community.

\end{abstract}


\IEEEpeerreviewmaketitle

\glsresetall

	\positiontextbox{11cm}{27cm}{\footnotesize \textcopyright 2024 IEEE. This version of the article has been submitted to IEEE Internet of Things Journal. Personal use of this material is permitted. Permission from IEEE must be obtained for all other uses, in any current or future media, including reprinting/republishing this material for advertising or promotional purposes, creating new collective works, for resale or redistribution to servers or lists, or reuse of any copyrighted component of this work in other works.}

\section{Introduction}

Nowadays, the concept of smart building is already a reality. A smart building is one that, based on the information obtained by a series of sensors, is capable of suggesting or directly making autonomous decisions that affect its internal functioning. Systems such as lighting, heating or cooling adapt in real time to the needs of people, allowing for a more efficient use of available resources. The concept of ``smart'' is also present in other areas, such as smart cities or smart homes, but in all of them the philosophy is similar: Capture information from the environment to provide it to the actuators so that they can operate efficiently.

There are a multitude of sensor types on the market, capable of measuring various physical parameters. Temperature, air quality, brightness, or power consumption meters are just a few examples. However, there is another type of sensor that is vital for a complete implementation of the ``smart'' concept in buildings: people monitoring sensors. This is a fundamental concept, since the number of people in the different rooms makes it possible to focus available resources on those areas of interest. Activating certain systems only in areas with people, or deactivating them in empty areas, are just some examples of decisions that can be made if the location of people is known in real time.

For most of the actions that can be carried out, it is not necessary to know the identity of each individual, but it is enough to know the total number of people in each area and how they move within the building. This fits perfectly with the growing emphasis on privacy that exists today, not only in this area but in many others. Users are increasingly demanding greater protection of their personal information, and are less willing to provide it to third parties.

As for technologies capable of providing the location data that would be needed in a smart building, there are many and varied. But from a practical point of view, they could be divided into two groups: those technologies that require each person to carry a locator element, and those in which this is not necessary. In the first group we could include the majority of systems based on radio frequency (UWB, Wi-Fi, Bluetooth, etc.), where each person is required to carry an electronic device that serves as a ``tag'', which is the entity that is actually detected. Thus, with these technologies, a person who does not carry a tag is totally invisible, but, on the other hand, by having a tag associated with each person, the system is able to know not only the position or distribution of people, but also the location of a specific individual.

There is, however, another family of technologies capable of providing location data without the need for people to carry any type of device, so the localization procedure becomes transparent to the user. These are systems such as those based on image capture, presence sensors, or radar technology. In these cases the sensing devices are able to perceive the disturbances caused by the human body in its environment, either by blocking or reflecting light, or by bouncing a radio signal. However, there are differences between them in terms of the risk to the privacy of users. Thus, techniques such as image recognition employ a type of measurement (images or frames from a video) that can be directly used to identify individuals or the activities in which they are participating. This poses a clear risk to the privacy of users, which can be compromised if such measurements are not effectively protected. In contrast, other technologies such as mmWave \gls{FMCW} radar employ a type of measurement (point clouds with estimations of distance, velocity, and angle between the target and the radar) that, in isolation, does not enable the individualized identification of a person or the discernment of the actions they are performing. Only after processing the data \cite{9780203}, and invariably requiring some form of additional external information, can radar measurements potentially be utilized to identify a person or their actions. This characteristic endows mmWave \gls{FMCW} radar technology with an additional layer of security in terms of safeguarding user privacy

The mmWave \gls{FMCW} radar technology employs the radar principle using short-wave electromagnetic signals in the millimeter range. Consequently, the frequency bands in which these devices operate are in the tens of gigahertz range, typically above 60--70 GHz. The basic idea of these systems is to emit an electromagnetic signal with certain characteristics and receive that signal back after it has bounced off the environment. Depending on the differences obtained between the two signals, the position of objects and people and their velocity can be inferred. By using an array of antennas, these devices can estimate not only these two parameters, but also the detection angle with respect to the radar.  

mmWave \gls{FMCW} radar devices have been used successfully for years, especially in the automotive sector \cite{8830483, 8827996, 9127853,9635901}. However, the technology can also be used in other fields such as the positioning of people in indoor environments. As for the problems and limitations of this technology in indoor environments, these are mainly due to the multipath effect and sparsity of the channel. The first phenomenon causes the receiver to detect the reflected signals from multiple paths after hitting different obstacles. This causes the appearance of erroneous detections, also called ``ghosts'' \cite{9266283,liu2020multipath}, that reduce the reliability of the system if they are not properly eliminated. On the other hand, although also related, the problem of sparsity refers to the low density of detection points. In practice, these systems use a \gls{CFAR} algorithm \cite{9581073} to detect the peaks in the different \glspl{FFT} performed after receiving the echoes of the transmitted signal. In indoor environments with multiple obstacles and reflections, the performance of this algorithm may decrease, reducing the number of samples. In addition, the usage of mitigation techniques for the problem of ghosting causes even more samples to be eliminated.

Despite these limitations, several works have presented systems based on \gls{FMCW} radar to perform localization tasks \cite{10017181,safa2023fmcw,8400812,8249147}. However, all of them have focused on the use of a single radar device. This is because the presence of several radar devices in the same area presents a challenge: the fusion of their data. Precisely because these systems do not use tags or other identifying elements, there is no direct way to ensure that the detections of the same person or object by several radars are indeed the same. But the use of a single radar makes it impossible to monitor large areas beyond the range of the device, and this is a major limitation for use cases such as indoor localization, where sometimes the areas of transit are too large to be covered by a single radar.

Thus, the main contribution of the present work is the proposal of a flexible software architecture to combine the information coming from several \gls{FMCW} radar devices to obtain a common position estimation of all targets present in an indoor space. To achieve this, several software modules were developed to perform the following tasks: read measurements from commercial \gls{FMCW} radar devices, combine different reference frames from each radar device into a single common frame, filter the samples to eliminate possible error-ghosts, perform clustering tasks to group the detected objects generated by the same target, track each of those targets continuously and, finally, generate useful information in a smart building environment (such as the number of people present in a given room) ready to be integrated in a common \gls{IOT} management platform such as Home Assistant \cite{home_assistant}. All the source code for these modules has been released as open source and is publicly available for download \cite{src_mmwave_reader, src_mmwave_cluster, src_mmwave_fuse, src_door_counter}.

In addition, as an example of use, a test was conducted in an indoor space using three mmWave \gls{FMCW} radar devices operating at the same time. As the ground-truth, video cameras and machine-learning-based people tracking algorithms were used. All collected data have also been published and it is available for the scientific community \cite{barral_vales_2024_10572015}.


The structure of this text is as follows. In \cref{sec:scenario}, we detail the psychical setup, including a description of the scenario, the hardware used, and the deployment process. In \cref{sec:software}, we specify the software architecture used, explaining each of the implemented layers, their operation, and parameters. Finally, in \cref{sec:test} we show the results of the example of use in the previous scenario.

\section{Scenario Setup}\label{sec:scenario}

In this section, we detail all the elements corresponding to the physical deployment. In \cref{subsec:mmwave}, we describe the \gls{FMCW} radar devices used, including basic notions of how the technology works. In \cref{subsec:deploy}, we describe the considered scenario and the deployment of the radars, indicating how they were placed and the necessary measures taken to unify their measurements.

\subsection{Description of the mmWave \gls{FMCW} Radar Devices Used}\label{subsec:mmwave}

In this work, we have used mmWave \gls{FMCW} radar devices from the manufacturer Texas Instruments (TI). We employed two different models, both based on the IWR6843 chip, which operates in the 60\,GHz to 64\,GHz frequency band and includes an ARM R4F-based processor, a \gls{RF} transceiver, and a \gls{DSP} for advanced signal processing. Both the \gls{MCU} and the \gls{DSP} can be programmed by the final user, hence different applications can be deployed in the same chip to take advantage of the mmWave \gls{FMCW} radar capabilities. 

Two different devices based on the IWR6843 were used in this work: the IWR6843ISK and the IRW6843AOPEVM. Both are evaluation boards that include an IWR6843 and several interfaces that allow for accessing and programming the on-chip \gls{MCU} and \gls{DSP}. The main difference between them lies in the type of their antenna array. Thus, the IWR6843ISK model mounts the array directly on the PCB, whereas the IWR6843AOPEVM has a special version of the IWR6843 (the IWR6843AOP) in which the array is integrated within the package. Also, there is a difference in the antenna configuration that results in the IWR6843AOPEVM model having a range of 120\textdegree\ of elevation \gls{FoV}, while the IWR6843ISK model has only 30\textdegree.

\subsection{Scenario and Deployment}\label{subsec:deploy}

\begin{figure}[!ht]
\includegraphics[width=1\columnwidth]{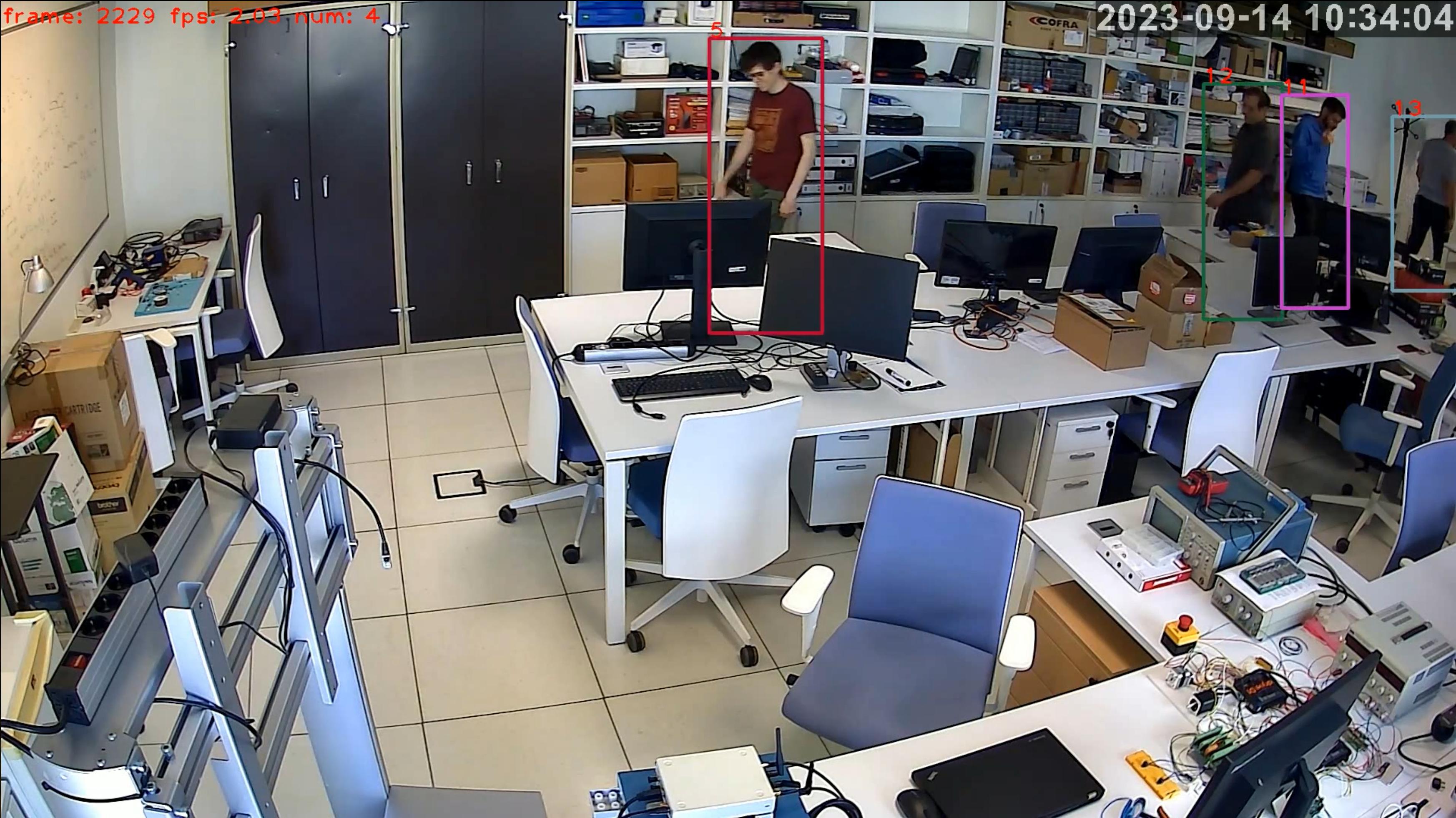}
\caption{Picture of the laboratory corresponding to a frame of one of the videos used to track people during the experiment.\label{fig:frame_video}}
\end{figure}

The scenario chosen for testing was one of the laboratories of the ``Scientific Area'' building of the University of A Coruña, Spain. The room has dimensions of 12$\times$6 meters, with a height to the ceiling of 2.35 meters. Inside are the different stations of the researchers, their computers, as well as many electronic devices of different nature. There are tables and chairs scattered throughout the space, and cabinets where books and cardboard boxes are stored. A photograph of this laboratory is shown in \cref{fig:frame_video}. 

\begin{figure}[!ht]
\includegraphics[width=1\columnwidth]{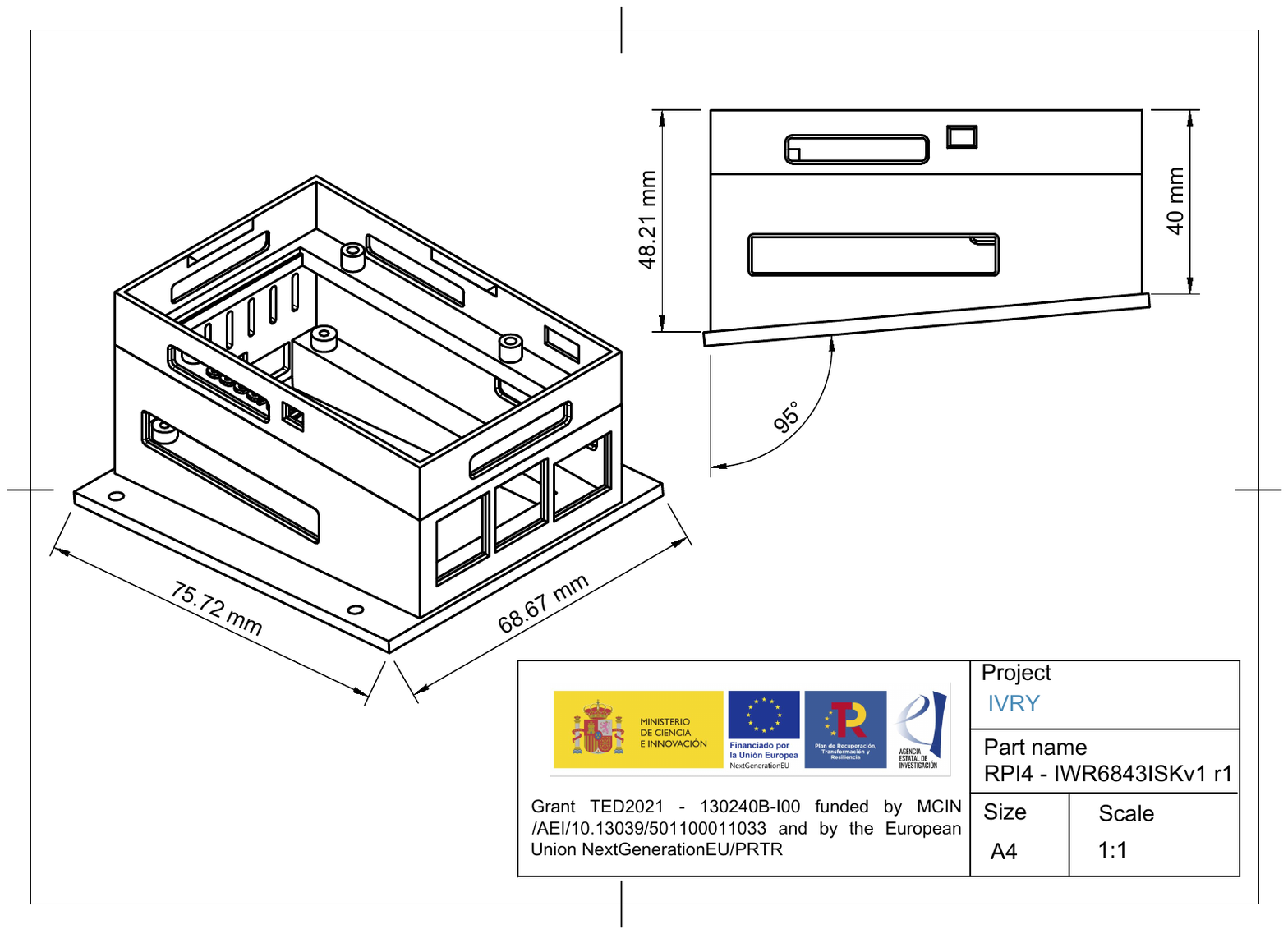}
\caption{Case designed to place the IWR6843ISK with a fixed angle on the Z-axis.\label{fig:3d_case}}
\end{figure}

Three \gls{FMCW} radar devices were installed in total. Two of them employ the IWR6843ISKl and are placed on the farthest walls of the room, and the other one, equipped with the IWR6843AOPEVM, is installed on the ceiling in the center of the room. One of the most important tasks when placing the devices is to accurately measure the position and orientation of each device. This is of vital importance since the estimates provided by the radar are always within the reference frame of the radar itself. To minimize the chances of error in taking measurements, we decided to build a 3D printed case so that the radar device is positioned with a preset inclination, as shown in \cref{fig:3d_case}.

\begin{figure}[!ht]
\includegraphics[width=1\columnwidth]{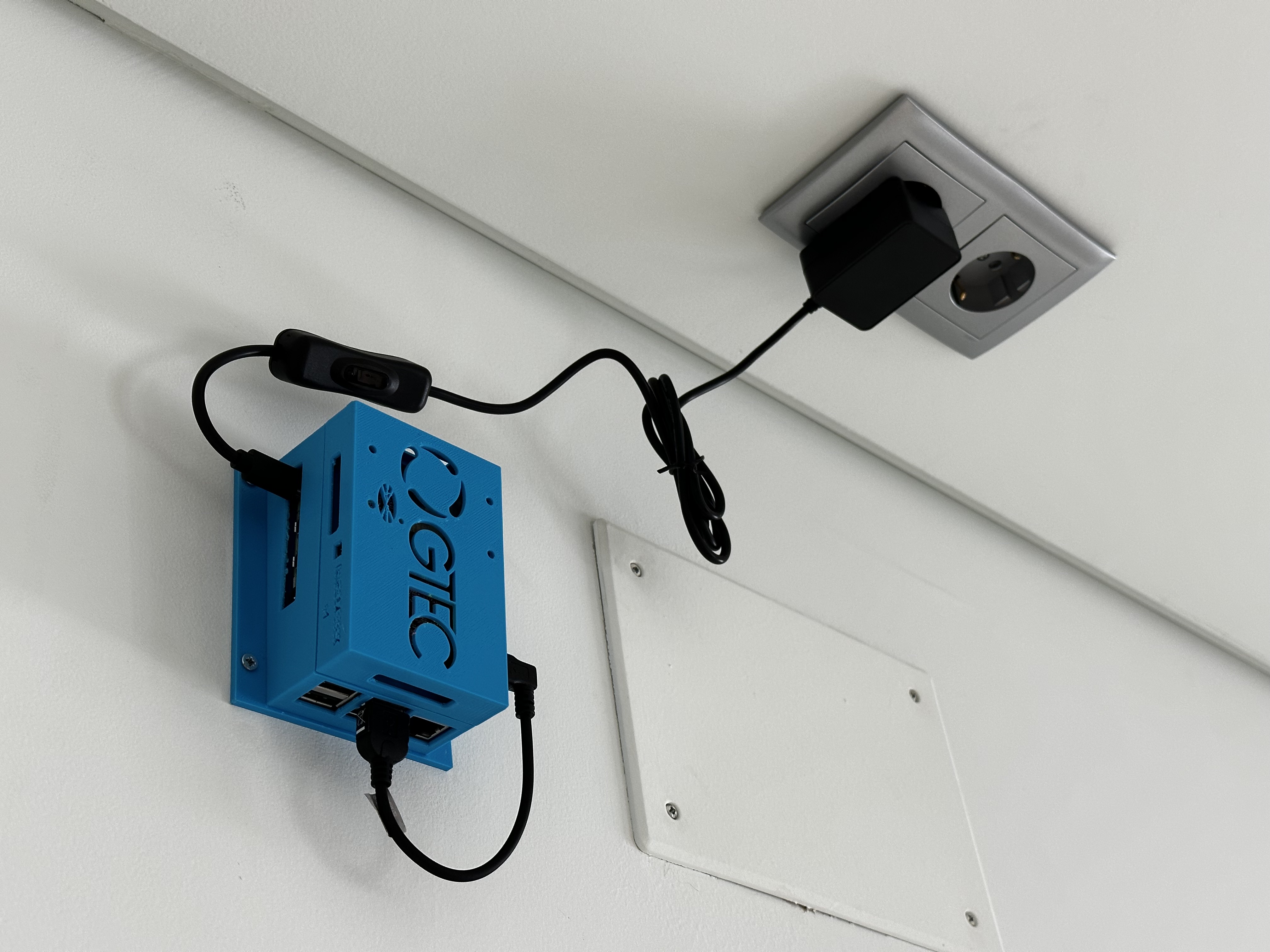}
\caption{Case with the IWR6843ISK module and the Raspberry Pi 4 placed on the wall.\label{fig:isk_place}}
\end{figure}

\begin{figure}[!ht]
\includegraphics[width=1\columnwidth]{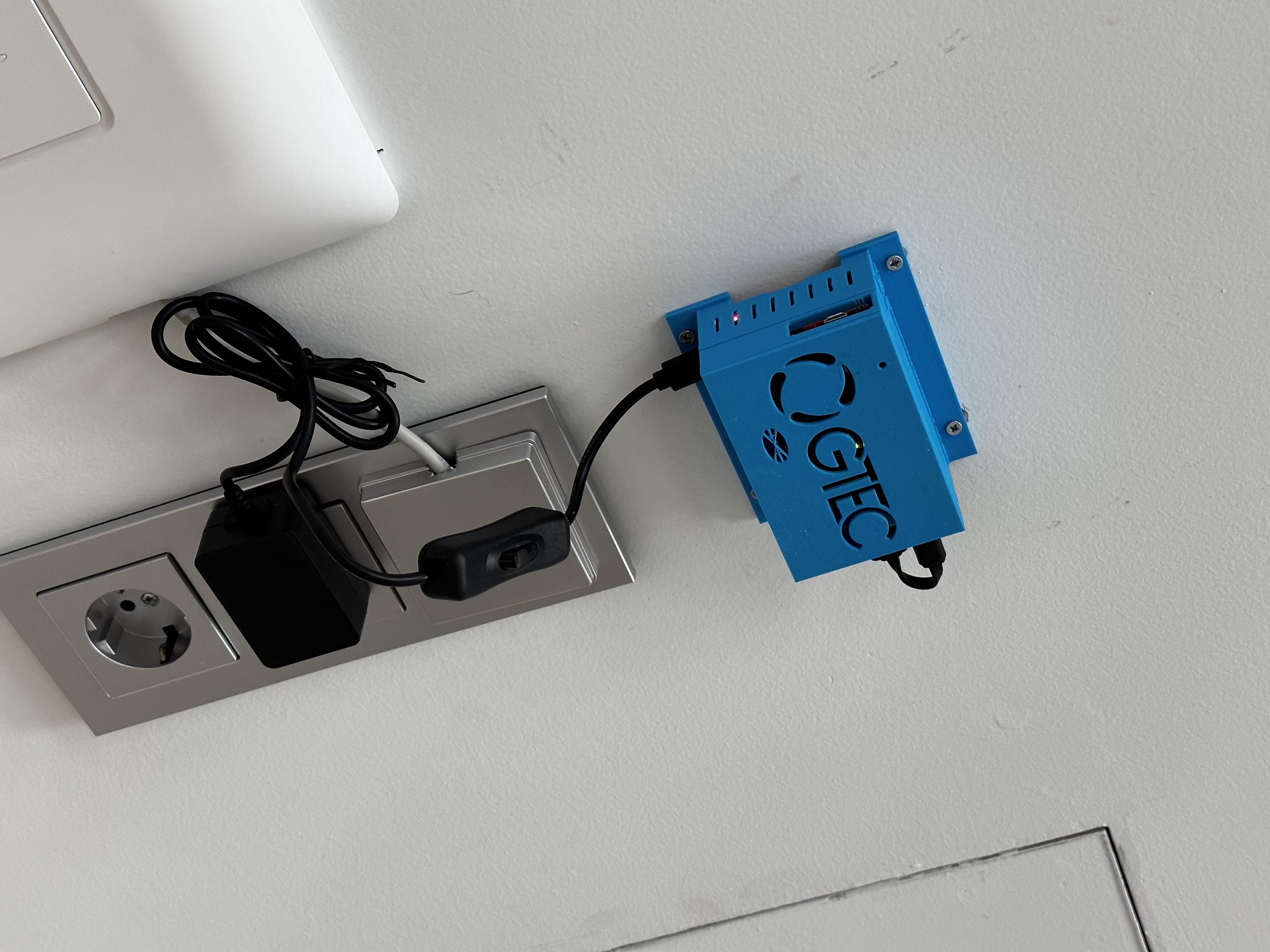}
\caption{Case with the IWR6843AOP module and the Raspberry Pi 4 placed on the ceiling.\label{fig:aop_place}}
\end{figure}

\cref{fig:isk_place,fig:aop_place} show the final placement of the IWR6843ISK (wall) and IWR6843AOPEVM (ceiling) models. Two different case versions were built, one with a tilt angle of 5° for the devices placed on the walls, and another one with an angle of 0° for the device placed on the ceiling. To accurately measure the final placement of the devices, different laser meters were used to obtain the X,Y, and Z values of each device with respect to the point chosen as the origin, which corresponds to one of the corners of the laboratory. 

\begin{figure}[ht]
\includegraphics[width=1\columnwidth]{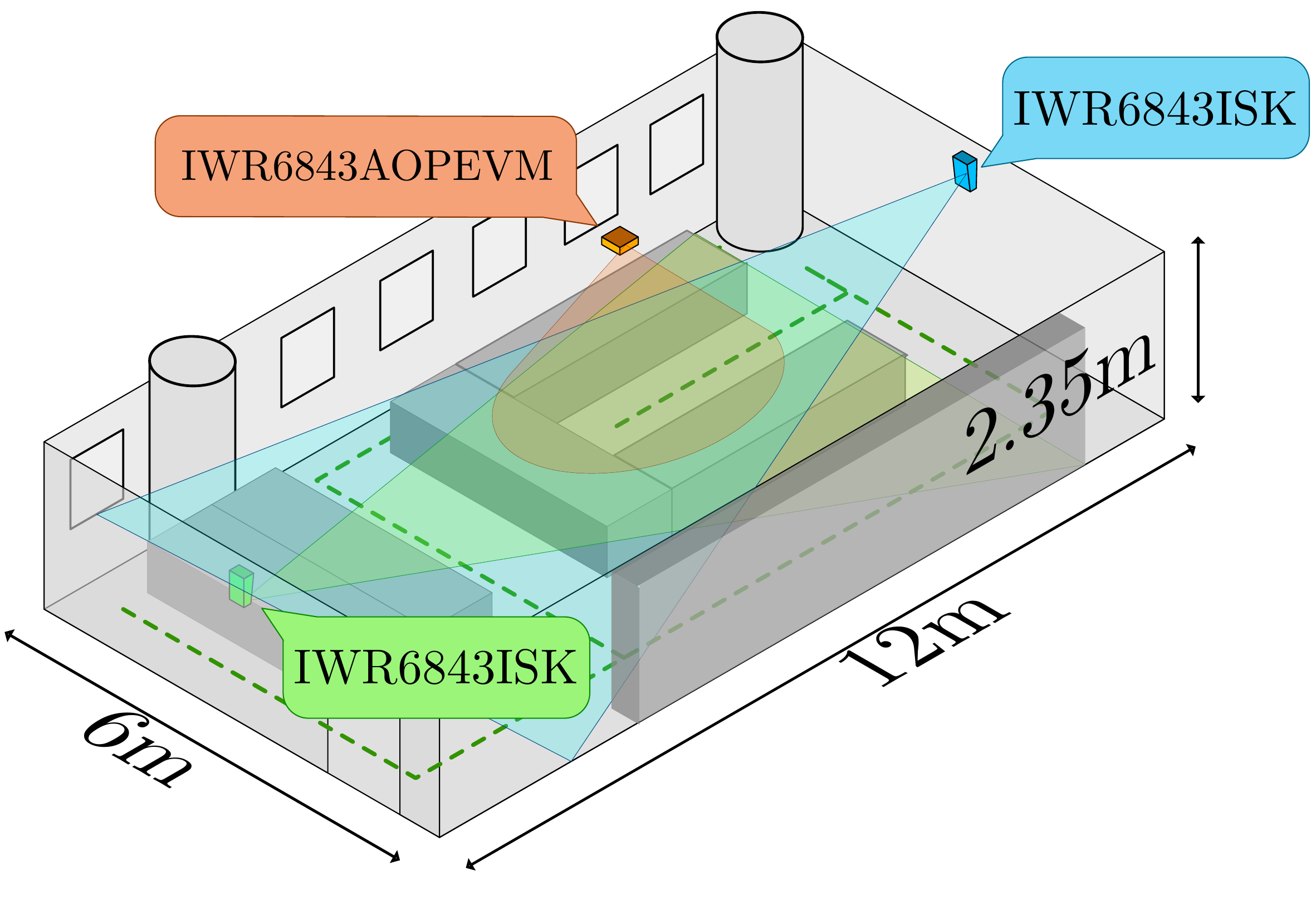}
\caption{Diagram of the laboratory with the radar devices in place.\label{fig:plan}}
\end{figure}

The three radar devices were positioned facing the center of the room. In choosing the radar signal parameters for each of them, the dimensions of the room were taken into account, so that the radars placed on the side walls could detect motion near the opposite wall. Because of the angle of inclination in the lateral axis (pitch), the area closest to each of these devices is a blind zone for each of these radars, and only the radar on the opposite wall is capable of detecting that zone. The radar placed on the ceiling was configured to maximize lateral detection (with respect to the radar), so that it could detect objects in the largest possible area under the device. \cref{fig:plan} shows a diagram with the placement of each radar and its approximate coverage area.

The position and orientation of each device was configured within a \gls{ROS} \cite{ros_web} environment. \gls{ROS} is an open-source framework designed for building, testing, and deploying robotic systems. It provides a collection of tools, libraries, and conventions that simplify the development of complex robotic systems by providing a standardized way of managing hardware abstraction, communication, and software distribution. Its architecture is based on a distributed messaging system, where different software components, called nodes, communicate with each other using messages. Nodes are independent processes that can run on the same computer or on different computers connected through a network, and they communicate with each other through a publish-subscribe mechanism, where a node publishes its data in a certain topic (a unique identifier within ROS) and other nodes subscribed to this topic receive it in real time.

Thus, the position an orientation of each radar was configured as a succession of transformations between the different frames (reference system) present in the test.  The idea is to apply a transformation between the local frames of each radar and a common global frame. To this end, it is necessary to rotate and translate each position depending on the orientation of each radar. This is why the radar positions and orientations must be measured accurately when they are physically placed in the room. To perform this task, the TF2 node, included in the \gls{ROS} distribution, was used. This node allows for performing the necessary frame transformations (configuring it with the position and rotation of each element), in a hierarchical way, hence enabling the transformation of the measurements so that all of them are in a common coordinate system.

In addition to the radars, two video cameras were deployed in the laboratory and placed on a tripod in two corners of the room. The purpose of these cameras was to obtain a video through which, using artificial vision and machine learning techniques, obtain the ground truth of the number and position of people inside the room during the test. The system used is explained in detail in \cref{sec:test}.

\section{Software Setup}\label{sec:software}

\subsection{Description of the Firmware Used}

By default, both IWR6843ISK and IWR6843\-AOPEVM devices do not come with any software loaded. However, TI provides the end user with a toolbox to work with IWR6843-based devices, the so-called Radar toolbox \cite{radar_toolbox}, which contains a collection of different applications and resources that allow the radar to be used in different scenarios. Among the different applications available, there are examples for working with radar in industrial, automotive, and domestic environments. In our case, to detect the movement of people within the individual range of each radar, we decided to use the following two similar example applications from the Radar toolbox: \textit{1)} \enquote{3D People Counting}, and \textit{2)} \enquote{Overhead 3D People Counting}. The fundamental difference between both applications is that one of them is optimized for using the radar device placed on a wall, whereas the other one is optimized to work when the radar is placed on the ceiling pointing directly to the ground. The applications can be divided into three main modules: \textit{1)} obtaining samples from the radar, \textit{2)} low-level processing of these samples to detect moving objects, and \textit{3)} a tracking phase that allows for successive detections to be associated with the same target (in this case, a person). Both radar models include detailed documentation, which allow to adapt them to the needs required by the end user. An official configuration guide is provided for the person detection module, and a separate guide is available for fine-tuning tracking parameters \cite{detection_tunning,tracking_tunning}.

The detection module is responsible for transforming the complex signals coming from the \gls{ADC} into measurement vectors known as point clouds. Each of these points represents a reflection point that has range, azimuth, elevation, radial velocity (i.e., Doppler), and \gls{SNR} values associated with it. The detection algorithm varies to suit each configuration depending on whether the radar device is placed vertically on a wall or horizontally on the ceiling.

The point cloud generated in the detection phase serves as input to the next module of the application: the tracking module, which groups the points and associate them to the different targets (people) present in the scene. To perform this tracking, an algorithm called group tracking is used. This algorithm is based on a variant of the \gls{EKF} where the usual phases of prediction and update are executed on groups of related samples. In the present work, this part of the implementation has been replaced by a series of \gls{ROS} nodes, as detailed in \cref{subsec:software_stack}. The reason for this is that the TI algorithm for performing tracking is limited to a single radar, so in a system with multiple radars it would be necessary to configure each radar differently and then merge the high-level positions. Another limitation of this algorithm is that it imposes a maximum limit of 8 simultaneous targets, which may be insufficient in some scenarios.

\subsection{Description of Captured Data}

Both the \enquote{3D People Counting} and the \enquote{Overhead 3D People Counting} applications allow for receiving in real time a series of measurements through the USB port. These measurements are encoded using the \gls{TLV} protocol, where each message includes an indication of its type, its length, and the corresponding value. The data structure of the header of these messages, with the type and length fields, can be seen in \cref{tlv_header}. 
\begin{lstlisting}[language=c, caption=Header structure., label=tlv_header, stringstyle=\ttfamily]
tlvHeaderStruct = struct(
    'type',    {'uint32', 4}, 
    'length',  {'uint32', 4});
\end{lstlisting}
The value of the measure depends on its type. Thus, both the \enquote{3D People Counting} and the \enquote{Overhead 3D People Counting} applications generate two main output types: \texttt{MSG\_COMPRESSED\_POINTS} and \texttt{MSG\_TRACKERPROC\_3D\_TARGET\_LIST}. The first one represents the set of points detected by the system after the detection phase. This set of points, called point cloud, is the basis on which the tracker algorithm works to follow each target. Each of the individual points is modeled according to the structure defined in \cref{tlv_points}. The fields included are:
\begin{itemize}
    \item $elevation$: Detection height with respect to the line perpendicular to the radar.
    \item $azimuth$: Angle in the horizontal plane between the detected point and the radar center.
    \item $doppler$: Estimated velocity with respect to the radar.
    \item $range$: Estimated distance from the radar to the point.
    \item $snr$: \gls{SNR} value.
\end{itemize}

The second type of structure, \texttt{MSG\_TRACKERPROC\_3D\_TARGET\_LIST}, is not used in this work, \todo{Explicar lo que hace cada tipo de dato, y limitaciones (8 targets, 1 radar, etc).} since the clustering and tracking phases are performed with an alternative implementation based on our \gls{ROS} nodes.

\todocheck[inline,color=pink]{Tomás: Los tipos de salida están definidas en el código de las aplicaciones? o dónde? Realmente es mejor mencionarlos o no? En caso de que sí, no sería mejor usar una fuente monoespaciada?, i.e., \texttt{MSG\_COMPRESSED\_POINTS} y  \texttt{MSG\_TRACKERPROC\_3D\_TARGET\_LIST}}{Yo creo que se pueden poner para que quede claro con qué tipo de datos trabajamos desde los radares. Quizás el mensaje con la estructura del header se puede quitar.}

\begin{lstlisting}[language=c, caption=Cloud point structure., label=tlv_points, stringstyle=\ttfamily]
pointStruct = struct(
    'elevation',  {'int8', 1},
    'azimuth',    {'int8', 1},
    'doppler',    {'int16', 1}, 
    'range',      {'uint16', 2},
    'snr',        {'uint16', 2});
\end{lstlisting}

\begin{figure}[!ht]
\includegraphics[width=1\columnwidth]{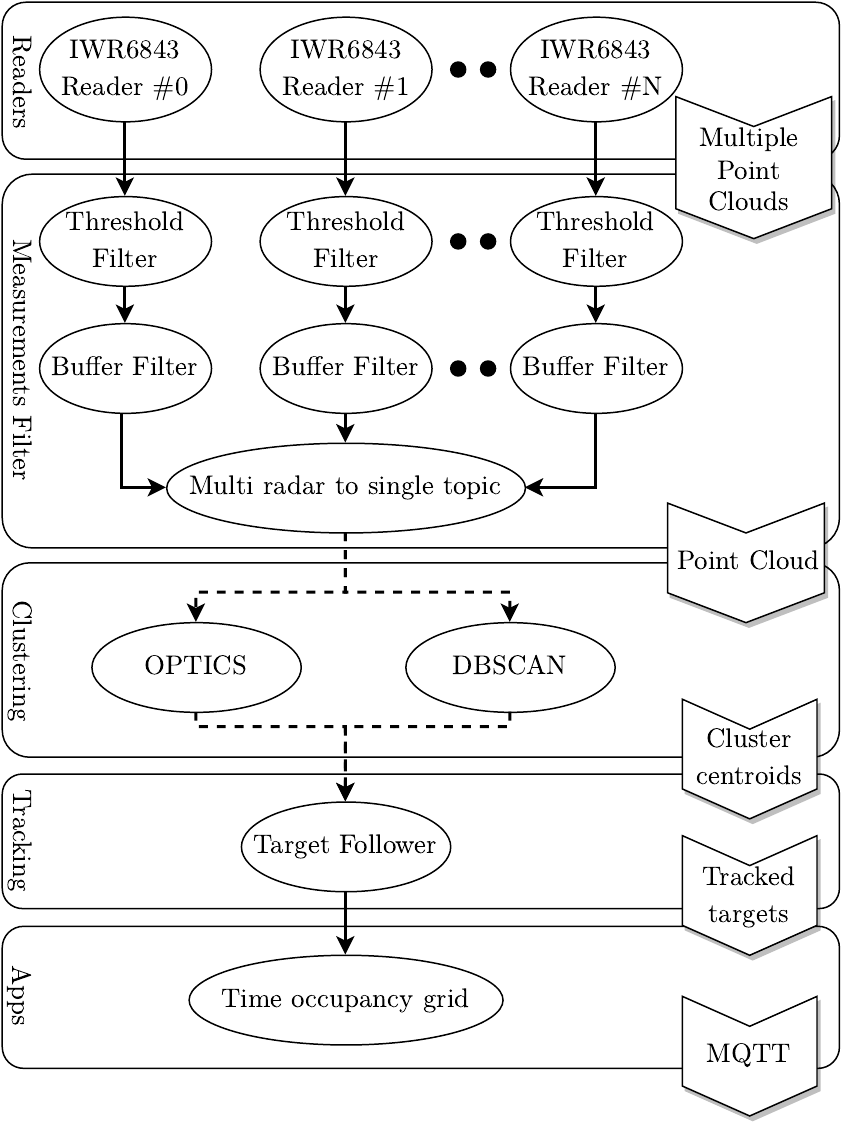}
\caption{ROS nodes architecture.\label{fig:ros_nodes}}
\end{figure}

\subsection{Multi-Radar Processing Stack\label{subsec:software_stack}}

\cref{fig:ros_nodes} shows the different layers of the proposed architecture to process and integrate measurements from several radars into a common reference frame. Each layer constitutes a different and self-contained abstraction level, hence modifications or improvements in one layer are transparent to the rest.

The first layer of the proposed architecture shown in \cref{fig:ros_nodes} corresponds to that of the reader nodes. This layer includes several instances (one for each radar device) of the \textit{IWR6843 Reader} node \cite{src_mmwave_reader}. The mission of this node is to connect, through the USB port, with the IWR6843ISK or IWR6843AOPEVM devices, send them a working configuration, and start receiving measurements. These measurements are next published in a \gls{ROS} topic to which the higher layer nodes can subscribe. The output of this layer will be point cloud messages, each consisting of multiple radar detection points, including ranging, azimuth, and elevation values.

The next layer of the proposed architecture is devoted to filter the radar measurements since they are subject to different types of errors that may cause their values to deviate from the physical reality present in the scenario. To mitigate these problems, the filtering layer consisting of several interconnected nodes is established. The first one, called \textit{Threshold Filter}\cite{src_mmwave_reader}, is a very basic filter that allows us to discard measurements with very low \gls{SNR} values or with abnormally high Doppler values. Thus, the node is configured with the desired thresholds and subscribes to the measurements coming from the reader, so that its output will be the samples that are within the values considered correct. The second node type is called \textit{Buffer Filter} \cite{src_mmwave_reader} and its mission is to eliminate the points that appear spontaneously, but disappear in successive measurements. In most cases, such detected values are erroneous, hence this node is responsible for eliminating them by applying an algorithm based on the continuity or not of contiguous measurements in time in areas close in space. Thus, the measurements that occur in an area of the space in which there are no more subsequent detected values are considered errors and are not published. 

Up to this level, filtering is performed individually for each radar. To integrate all the measurements in a single common topic, the \textit{Multi-radar to single topic} \cite{src_mmwave_fuse} node is used, whose only mission is to subscribe to the output topics of the filters of each radar and to republish all the measurements together in a new common topic. Upon exiting this layer, consequently, a set of measurements will be established, common to all the available radars, with each measurement no longer retaining a direct association with the specific device that captured it.

The next layer is the clustering layer, where different nodes \cite{src_mmwave_cluster} are used to identify the cloud points that correspond to the same physical object. Thus, for each set of points captured within a specific temporal window, the nodes corresponding to this clustering layer will generate a set of clusters based on various metrics, such as point density or the distance between points. To conduct the experiments, this work has implemented two well known blind clustering algorithms extensively documented in the literature: \gls{OPTICS} \cite{ankerst1999optics} and \gls{DBSCAN} \cite{ester1996density,schubert2017dbscan}. Both \gls{OPTICS} and \gls{DBSCAN} are density-based spatial clustering algorithms aiming to identify clusters based on their local density variations. Both algorithms begin in a similar way, by selecting an arbitrary data point and calculating its distance (according to a given metric) to its neighboring points within a specified radius. This distance represents the degree of density at which the point is reachable. The process then continues to expand the cluster by examining the reachability of the neighboring points and adding them to the cluster if they meet certain density criteria. The primary distinction between these two algorithms lies in their approach to cluster density. \gls{DBSCAN} assumes a uniform density across all clusters, whereas \gls{OPTICS} is designed to accommodate varying densities. Although these two algorithms were implemented in this work as an example, the proposed architecture is so flexible that it would support additional nodes to perform the clustering task without affecting the rest of the nodes of the other layers. To do so, it would simply be necessary to take into account the type of measurements to be received (the global point cloud, resulting from merging the points of the different radars), and the type of measurements to be published as a result which, in this case, is the set of centroids of the detected clusters.
 
After the \textit{Clustering} stage, in the proposed architecture the data flow reaches a new stage called \textit{Tracking}. The objective of this stage is to perform a dynamic tracking of the cluster centroids identified in previous stages, so that the final result is a position estimation that takes into account not only the positions detected at a given instant by the radars but also their dynamics (i.e., velocity and acceleration). The tracking layer is implemented by means of a new node named  \textit{Target Follower} \cite{src_door_counter}, whose operation is based on the use of an \gls{EKF} that is dynamically created each time a new target is identified in the clustering phase. However, before generating a new instance of an \gls{EKF} to follow a target, the following checks are performed:
\begin{itemize}
    \item Upon receiving a new centroid from the clustering layer, its compatibility is verified with the next estimate of an already running Extended Kalman Filter (\gls{EKF}).
    \item If it is compatible, then the position is considered a new measurement to feed the corresponding \gls{EKF}.
    \item  Otherwise, the position is treated as indicative of a new target in the system, prompting the creation of a new \gls{EKF} to monitor its motion.
    \item If a target remains without compatible measurements for an extended period, it is deleted along with its corresponding \gls{EKF}.
\end{itemize}
Both time and distance thresholds are configuration parameters of the node, together with the maximum number of targets to track simultaneously or the size of the temporal window to be applied.

At the end of the tracking stage, the system has a list of possible targets whose positions are generated based on the measurements captured by any radar in the scenario. Thus, high-level applications can now be built on this data. As an example, the proposed architecture implements the so-called \textit{Apps} layer including a node to count the number of people inside a room in real time. This node, termed as \textit{Time occupancy grid} \cite{src_door_counter}, allows for segmenting a physical space into logical zones of interest, hence making possible to have a real-time indication of which targets are in which zone and for how long. These zones of interest are defined by their center coordinate in the $XY$ plane, and two lengths, one in each of these two axes. The zones of interest can overlap with each other. In addition to the zone definition, the \textit{Time occupancy grid} node divides all the available space into a grid of cells (the dimensions of these are configurable). These cells have a hysteresis mechanism, so that the detection of occupation by a target is not immediate. Certain conditions (also configurable by the node) are needed to consider that a cell has been occupied (or unoccupied). This makes the target detection system within zones of interest more robust, at the cost of increasing the latency in the notification of state changes. The node publishes all the state changes that occur, either because a target enters a specific area of interest or when departs from it. Furthermore, at a higher periodicity, the current state of each zone is also published, i.e., which targets are within its boundaries and how long they have been inside.

\begin{figure}[!ht]
\includegraphics[width=1\columnwidth]{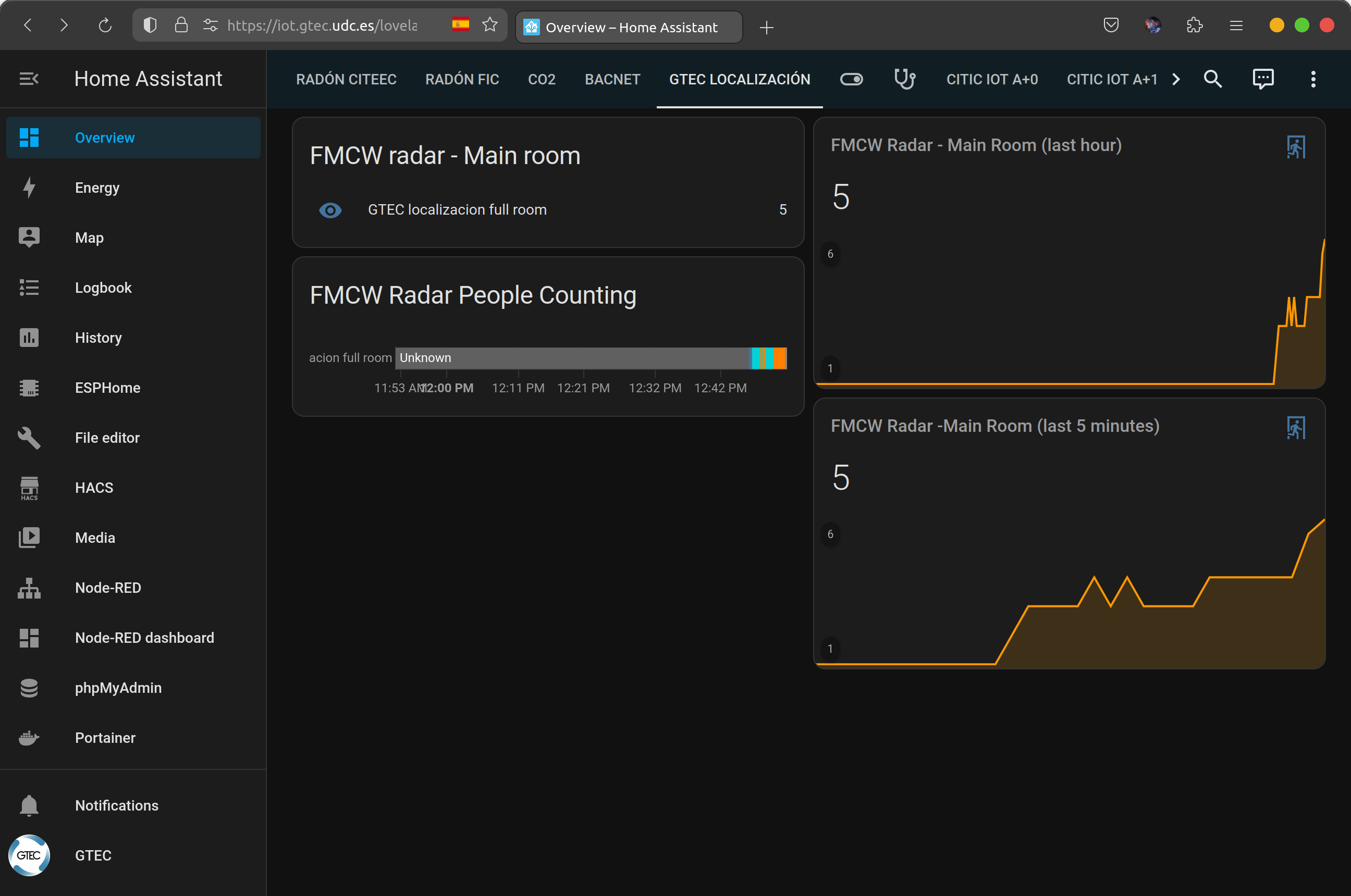}
\caption{\gls{IOT} platform receiving data generated by the system based on \gls{FMCW} radar devices. \label{fig:iot_dashboard}}
\end{figure}

To connect with the \gls{IOT} world, the \textit{Time occupancy grid} node allows for publishing its data via the \gls{MQTT} protocol, taking advantage of the deployment of an \gls{IOT} architecture at the University of A Coruña, carried out by the authors of this work \cite{dominguez2022overview}, and leading to a successful integration of the generated occupancy. \cref{fig:iot_dashboard} shows a screenshot of the Home Assistant instance deployed in that project, in which a graph shows the occupancy estimated by the deployed radar system. 

\section{Real Environment Proof of Concept}\label{sec:test}

As an example to validate the counting system with multiple simultaneous radars, a test was performed in a real environment. This section describes the experiment carried out and the results obtained.

The test consisted in obtaining measurements from the mmWave \gls{FMCW} radars and camera devices while a number of people entered the test room and moved around. The test began with a single person inside the room, and then three other people entered sequentially every 15 seconds. Once inside, people moved around in the room, passing through the coverage areas of the different radar devices. There were also moments when one or more of the people remained still, simulating the usual work sitting in front of a computer. At certain points during the test, several people joined together in a small space and then moved off again following different directions.

\begin{figure}[!ht]
\includegraphics[width=1\columnwidth]{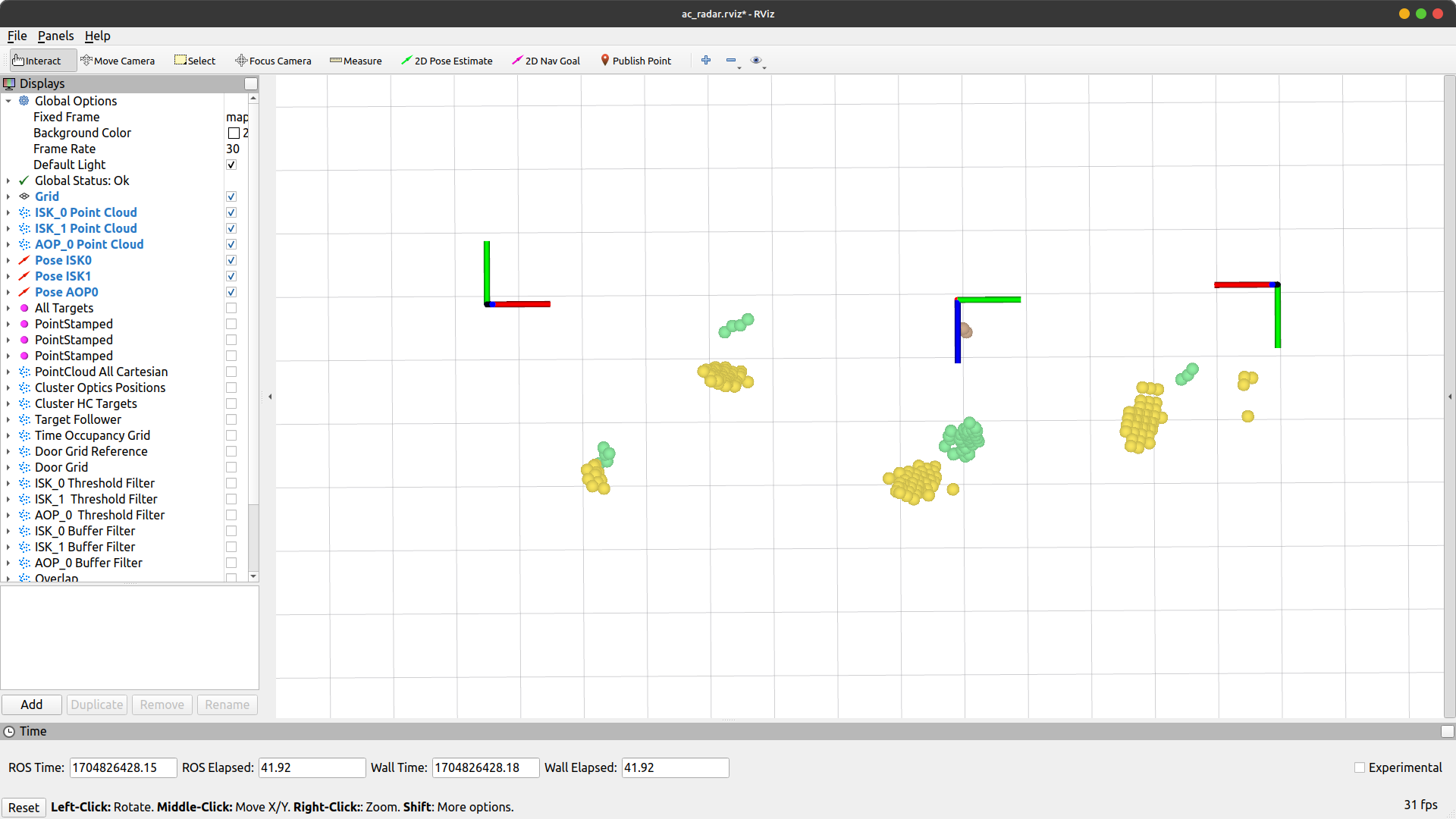}
\caption{RVIZ viewer showing some radar cloud points. \label{fig:rviz}}
\end{figure}

Leveraging the fact that the entire ecosystem is implemented in \gls{ROS}, the testing phase, as well as the post-testing period, enabled the utilization of various tools available in this environment, such as RVIZ, which is a 3D visualization tool for \gls{ROS} that facilitates the graphical representation of the different entities published by each node, thereby allowing for the visual observation of the effects on the outcomes (such as filtering, clustering, tracking, etc.) resulting from the modification and adjustment of various parameters. \cref{fig:rviz} shows a screenshot of this utility, displaying points detected by the three radars as several individuals moved around in the laboratory.

Another \gls{ROS} environment utility that was very useful during the test is Rosbag, which records messages and plays them back and allows for performing the test in two stages: a first phase for capturing a log with the raw measurements, both from the radars and the cameras, and a second phase, offline, in which the log is re-executed under different configurations or conditions. Therefore, it was possible to test the system, for example, with two different clustering algorithms, DBSCAN and OPTICS, simply by replaying the log while activating one node or another.

After the test, the positioning data provided by the multi-radar system were obtained and compared with those generated by the cameras. The videos were processed by the multi-camera person location and tracking system proposed in \cite{CARROLAGOA2023100940}, following the same processing described in the article and considering the same lightweight \gls{DNN} models. Person detection was performed using the PoseNet pose-estimation model \cite{posenet2018} with a ResNet50 backbone and a resolution of 640$\times$480. Person re-identification was performed using a MobileNetv2-based model provided by the Torchreid library \cite{torchreid}. The processing of the videos was performed at 2 frames per second. The only changes to the configuration detailed in \cite{CARROLAGOA2023100940} are the following:
\begin{itemize}
    \item The detections that were too far away from the cameras were discarded to avoid using detections with very low precision.
    \item The tracker's parameters were adjusted to enforce the tracks to start and end at positions next to the door, or at the beginning or the end of the videos.
\end{itemize}

\begin{figure}[!ht]
\includegraphics[width=1\columnwidth]{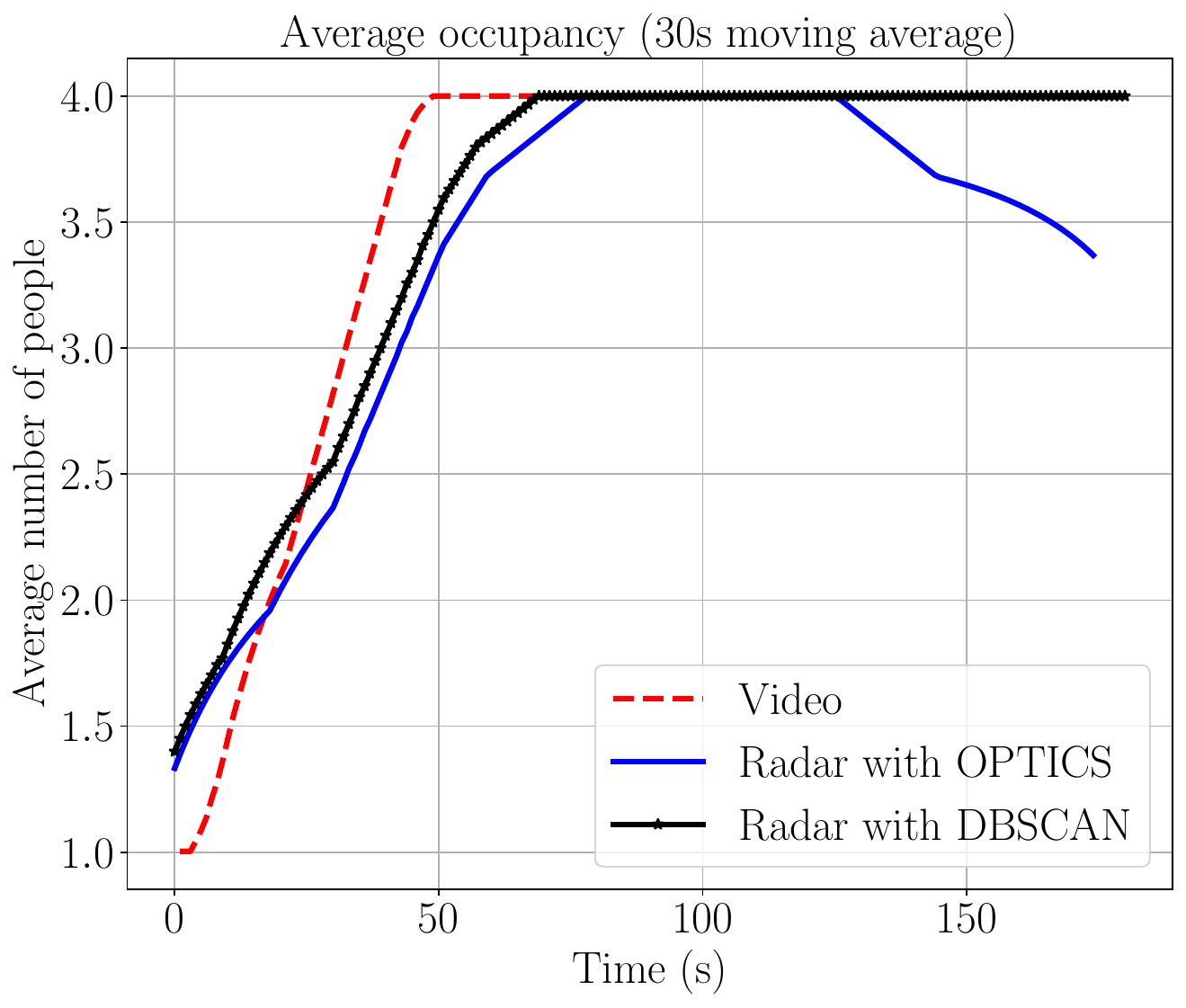}
\caption{Average people count using image tracking and three mmWave \gls{FMCW} radar devices combined.\label{fig:result_full}}
\end{figure}

As a result, \cref{fig:result_full} shows an average of the cumulative value of people detected by the radar-based and by the image-based systems. The values in this graph are averaged using a 30-second moving average. For the radar values, there are two different results: one using OPTICS algorithm in the clustering layer, and another one considering DBSCAN. We can see how the system based on multiple radars performs similarly to the reference system, although it needs some time to converge to the actual number of people. This is due to the different methods employed in the different nodes to try to eliminate false positives and ghosts, which means that it does not generate a positive detection until there are solid indications that it is indeed a person moving in the room.

Although the primary objective of the test was not to assess the performance of the system, it is clear that certain discrepancies emerge in the results when utilizing one clustering mechanism as opposed to another. Thus, the results show that with OPTICS clustering there is an underestimation of the number of people towards the end of the test because the algorithm does not find clusters that meet the density requirements during that time period, hence in the upper layers, the tracking algorithm does not receive new positions. According to the radar settings, only targets exhibiting movement were reported. Consequently, in the absence of motion, almost no point clouds are generated by the radar, and this lack of information results in the system relying on an ``inertia'' mechanism, wherein targets continue to be considered present despite the absence of detected movement over a time interval. If this motionless interval extends for a long time, the multi-radar system ultimately removes the target from the list of detected elements, deeming it a potential error. 

The parameters governing this behavior are distributed among the various nodes of the architecture, allowing for the system response to be tailored to each specific deployment scenario. Hence, the system necessitates an initial configuration based on the expected density of individuals in the area, their level of movement, and their typical periods of inactivity.

On the other hand, the example also served to verify the connection with an \gls{IOT} platform. To this end, the deployment developed during a previous study\cite{dominguez2022overview} was utilized. During the course of such a work, an instance of Home Assistant was deployed, capable of receiving, storing, and displaying data from various sensors positioned in different rooms in buildings at the University of A Coruña. Considering this setup, a virtual sensor was registered on this platform, which specifically provided occupancy estimates derived from the mmWave \gls{FMCW} multi-radar system. The integration was almost instantaneous upon the publication of the occupancy data in \gls{MQTT} format, necessitating only the configuration of the responsible node with the server address where the Home Assistant instance was operational. Post-integration, it was verified that the data were indeed being received and could be accessed using one of the multiple visualization tools available in Home Assistant, as shown in \cref{fig:iot_dashboard}.

\section{Conclusions and Future Work}
\label{sec:conclusions}

In this study, we introduced a software architecture designed to aggregate measurements from multiple mmWave \gls{FMCW} radar devices within the same indoor environment, creating a passive people counting system that respects user privacy by eliminating the need for individuals to carry electronic devices. To test the said software architecture, a real experiment was designed where three \gls{FMCW} mmWave radar devices were placed in the same room, accessed by different individuals. To verify the accuracy of the estimations, two cameras were simultaneously used to generate a ground truth based on image tracking. All the developed source code, along with the radar measurements and image processing, were published under an open source license \cite{src_mmwave_reader, src_mmwave_cluster, src_mmwave_fuse, src_door_counter, barral_vales_2024_10572015}.

The primary challenges encountered were linked to the inherent characteristics of radar measurements. Utilizing TI's software embedded in the ARM module of the radar devices, the obtained measurements were calculated using their \gls{CFAR} algorithm, resulting in numerous outliers in indoor scenarios. Additionally, due to bandwidth constraints when extracting measurements through the USB port, the data acquired from the radar devices represented only a subset of the generated data. For instance, with the experiment's configuration, no information was received when Doppler approached zero, indicating no movement. This limitation complicates maintaining an accurate people count when individuals cease movement and transition to a static state.

In future research, our primary focus is to address this issue, potentially by integrating radar devices with distinct configurations. One of these configurations aims to capture null Doppler values precisely, enhancing the system's ability to count people accurately, even when they are in a stationary state.

\bibliography{IEEEabrv,main} 
\bibliographystyle{IEEEtran}

\vskip -2.4\baselineskip plus -1fil
\begin{IEEEbiography}
	[{\includegraphics[width=1in,height=1.25in,clip,keepaspectratio]{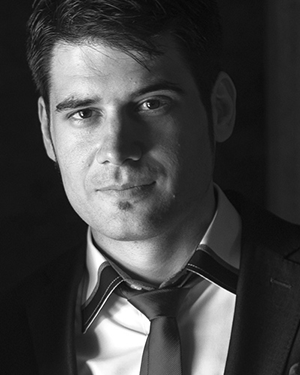}}
	]{Valentín Barral Vales} works as an associated researcher at University of A Coruña, where he took his degree as a computer engineer and his Ph.D. with a thesis titled: \textit{Ultra Wideband location in scenarios without clear line of sight}. His research field is indoor localization using radio technologies, a field in which he has published several works since 2012. He has also participated in many national and European projects, always related to indoor localization in complex environments.
\end{IEEEbiography}
\vskip -2.4\baselineskip plus -1fil
\begin{IEEEbiography}[{\includegraphics[width=1in,height=1.25in,clip,keepaspectratio]{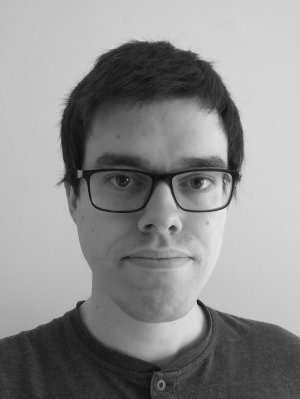}}
	]{Tomás Domínguez-Bolaño} received the B.S degree in Computer Engineering and the Ph.D. in Computer Engineering (with the distinction ``Doctor with European Mention'') from the University of A Coruña, A Coruña, Spain, in 2014 and 2018, respectively. Since 2014 he has been with the Group of Electronics Technology and Communications. In 2018 he was a Visiting Scholar with Tongji University, Shanghai, China. He is an author of more than 15 papers in peer-reviewed international journals and conferences. He was awarded with a predoctoral grant and two research-stay grants. His research interests include channel measurements, parameter estimation and modeling and experimental evaluation of wireless communication systems.
\end{IEEEbiography}
\vskip -2.4\baselineskip plus -1fil
\begin{IEEEbiography}
	[{\includegraphics[width=1in,height=1.25in,clip,keepaspectratio]{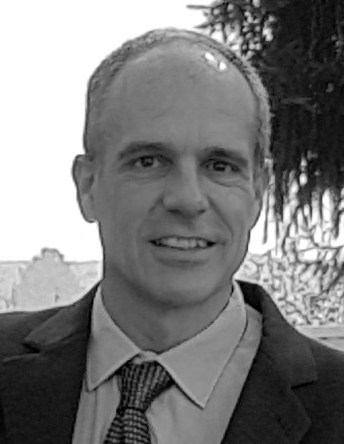}}
	]{Carlos J. Escudero} (Senior Member, IEEE) received the Ph.D. degree in Computer Science from University of A Coruña (UDC), Spain in 1998. He is a Full Professor at the University of A Coruña (UDC), Spain, in the area of Signal Theory and Communications. Since 1992, he has been a Faculty Member with the Department of Computer Engineering (UDC). He was Secretary of the Department, Vice-Dean of the Faculty of Informatics, Head of the Department of Electronics and Systems  and, since January 2016 to present, Deputy Rector of the UDC for Information and Communication Technologies (University CIO). His research focuses on applications of signal processing in communications, wireless sensor networks, and indoor location systems. Proof of this are some of the merits alleged in the curriculum: publications in the most prestigious journals and books, many national and international collaborations with renowned researchers, dozens of papers / publications in national and international conferences, highlighting the most significant in the curriculum, participation in more than 25 competitive research projects in which he has been the principal investigator on 7, responsible for 20 contracts with companies, author of two patents, and founding member of two spin-off companies.
\end{IEEEbiography}
\vskip -2.4\baselineskip plus -1fil
\begin{IEEEbiography}[
	{\includegraphics[width=1in,height=1.25in,clip,keepaspectratio]{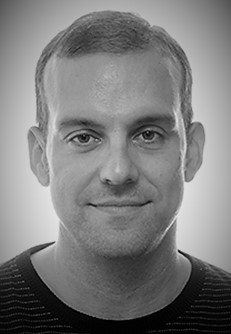}}
	]{José A. García-Naya} (Senior Member, IEEE) received the M.Sc. and Ph.D. degrees in computer engineering from the University of A Coruña (UDC), Spain, in 2005 and 2011, respectively. He is with the CITIC Research Center and with the Group of Electronics Technology and Communications, both at UDC, where he is Associate Professor. He is the coauthor of more than 120 peer-reviewed papers in journals and conferences. He was member of the research team in more than 40 research projects funded by public organizations and private companies, being principal investigator in two of them. His research interests focus on wireless engineering, with special emphasis on experimental evaluation, including wireless channel characterization, high-mobility vehicular transportation, time-modulated arrays; location systems, and IoT.
\end{IEEEbiography}

\end{document}